# Holographic Dark Energy from Acceleration of Particle Horizon

H. R. Fazlollahi*

*Institute of Gravitation and Cosmology, Peoples Friendship University of Russia (RUDN University), 6 Miklukho-Maklaya St, Moscow, 117198, Russian Federation*

**Abstract**

Following the holographic principle, which suggests that the energy density of dark energy may be proportional inversely to the area of the event horizon of the Universe, we have proposed a new energy density of dark energy through the acceleration of particle horizon scaled by the length of this parameter. This model depends only on one free parameter $\beta \approx 0 - 1.99$. For $\beta$s near zero, the deviation of the model compared with the $\Lambda$CDM model is tangible while for values $\beta \rightarrow 1.99$, the suggested model has no conflict with $\Lambda$CDM theory. Regardless of the value of $\beta$, the model presents dark energy that behaves such as matter with positive pressure in high redshifts, $\omega_X \approx 0.33$, while for present and near-future Universe treats like the cosmological constant model and phantom field. Comparing the model with Ricci dark energy illustrates our model alleviates Ricci dark energy errors in calculating the age of old supernovae and the evolution of different cosmic components in high redshifts. Moreover, we have calculated matter structure formation parameters such as CMB temperature and matter power spectrum of the model to consider the effects of matter-like dark energy during the matter-dominated era.

PACS numbers: 98.65.Dx, 98.80.Cq

## I. INTRODUCTION

Recent celestial observations like supernovae, CMB anisotropy, and galaxy clustering have strongly indicated that our Universe is spatial flat [1], and there exists an exotic cosmic fluid called dark energy with negative pressure that constitutes about two-thirds of the total energy of the whole Universe [2, 3]. This cosmological component is characterized by its equation of state $w$, which lies very close to $-1$, or according to some current data below $-1$ [4].

Many candidates including the cosmological constant, quintessence, phantom, quintom, holographic dark energy, etc. have been proposed to explain this mysterious component and its effects on the expansion of the Universe. However, people still do not understand what dark energy is. Ever since the discovery of dark energy, cosmologists are confronted with two fundamental problems: (1) the coincidence problem and (2) the fine-tuning problem [5]. The coincidence problem states the density of the dark energy and matter evolves differently while the Universe expands and now they are comparable, this is an incredibly great coincidence if there is not some internal connection between them. The fine-tuning problem traces back to the simplest form of the dark energy model, the cosmological constant model introduced by Einstein in which although vacuum energy in quantum field theory has the same property, it is greater than the observed value by some 123 orders of magnitude, and so we need extreme fine-tuning of the vacuum energy. Both of these problems are solved through different models for dark energy (see e.g. [6]).

Studying the black hole theory and string theory gives us new impressive sight to resolve dark energy problems in the new way called holographic principle, which may provide some clues about dark energy properties. It is realized that the entropy of a system scales not with its volume, but with its surface area in quantum gravity mediums [7]. Investigating Einstein's equation, $G_{\mu\nu} = 8\pi T_{\mu\nu} + \Lambda g_{\mu\nu}$, shows that the cosmological constant $\Lambda$ is the inverse of some length squared, $[\Lambda] \sim l^{-2}$. So according to this fact, in [8], it is proposed that unknown vacuum energy could be present as proportional to the Hubble scale $l_H \sim H^{-1}$. Consequently, the fine-tuning problem is solved by scaling dark energy through cosmological scale not by Planck length, and the coincidence problem is also alleviated. However, this model cannot explain the accelerated expansion of the Universe in late time wherein the effective equation of state for such vacuum energy is zero and the Universe is decelerating. Granda and Oliveros have proposed another possibility: the length $l$ is given by the average radius of the term $(\alpha H^2 + \beta \dot{H})^{-1/2}$ in which $\alpha$ and $\beta$ are some constants of the model. Consequently, in their model, the energy density of dark energy $\rho_X \propto \alpha H^2 + \beta \dot{H}$ which implies that the density of dark energy is proportional to the second Friedmann equation [9]. Different studies on this model are considered in [10]. Recently, C. Gao et al have proposed a new version of dark energy which satisfies the holographic principle. The length $l$ is given by the average radius of Ricci scalar curvature, $R^{-1/2}$, thus one could have the dark energy density $\rho_X \propto R$. This model presents the dark energy model in two terms, one part evolves like non-relativistic matter $\sim a^{-3}$, and another part which is slowly increasing with decreasing redshift [11]. This holographic dark energy model and its interacting versions are successful in fitting the current observations (see e.g. [12]). However, as shown in [11], the Ricci holographic dark energy as one of the explicit cases of the Granda-Oliveros dark energy model has some inconsistencies with the ages of some old objects. This problem was alleviated by considering Ricci dark energy as a viscous field [13].

---

* seyed.hr.fazlollahi@gmail.com
  Full name of author is seyed hamidreza fazlollahi.

In the other case, the particle horizon size $l_{PH} = a\int_0^t dt/a$ could be used as the length scale [14], but as shown in [15], the equation of state for this dark energy model is greater than $-1/3$, and so it still could not describe the late-time accelerated expansion. In this paper, we have concentrated on the particle horizon to show how this parameter yields a new version of holographic dark energy. Observations demonstrate the Universe expands with the acceleration phase in late time and so the acceleration of the particle horizon evolves proportionally to this acceleration rate.

$$\ddot{l}_{PH} = \dot{H}l_{PH} + H^2 l_{PH} + H \quad (1)$$

Checking acceleration of particle horizon shows $[\ddot{l}_{PH}] \sim l^{-1}$. As a consequence, rescaling (1) by the particle horizon $l_{PH}$ obtains $[\ddot{l}_{PH}/l_{PH}] \sim l^{-2}$. Hence, unknown vacuum energy, the energy density of dark energy could be present as proportional to the $\ddot{l}_{PH}/l_{PH}$.

In the next section, we describe this model and its cosmic equations. Sec. III is devoted to its structure formation and CMB anisotropy. In Section IV, we summarize the results of the model.

Throughout this paper, we adopt the Planck units, i.e. $c = \hbar = G = 1$.

## II. HOLOGRAPHIC DARK ENERGY

In this paper, we assume that the Universe is homogenous and isotropic and described by the Friedmann-Robertson-Walker metric

$$ds^2 = -dt^2 + a(t)^2 \left( \frac{dr^2}{1 - kr^2} + r^2 d\theta^2 + r^2 \sin^2\theta d\phi^2 \right), \quad (2)$$

where $k = 1, 0,$ and $-1$ describe closed, flat, and open Universe, respectively. The different independent observations confirm observable Universe is flat spatially, $k = 0$, and so the first Friedmann equation given by [1]

$$H^2 = \frac{8\pi}{3}(\rho_r + \rho_m + \rho_X) \quad (3)$$

where $H \equiv \dot{a}/a$ is Hubble parameter, over dot denotes the derivative with respect to the cosmic time $t$ while $\rho_r$, $\rho_m$ and $\rho_X$ are energy density of radiation, non-relativistic matter, and dark energy, respectively.

Following the holographic principle, we propose a new dark energy model in which the future event horizon area is proportional to the inverse of $\ddot{l}_{PH}/l_{PH}$. Therefore, the energy density of dark energy given as

$$\rho_X = \frac{3\alpha}{8\pi}\left(\frac{\ddot{l}_{PH}}{l_{PH}}\right) \quad (4)$$

where $\alpha$ is the constant of the model and the factor $3/8\pi$ is used for convenience in the following calculations.

To study a general holographic description, it is worthwhile we assume the conventional formula for the Hubble rate as [16]

$$H = \frac{\beta}{l} \quad (5)$$

where $\beta$ is a positive constant for an expanding Universe and $l$ denotes the cosmological length scale. For this scale, we use the particle horizon $l_{PH}$, and thus, Eq. (5) obtains particle horizon as a function of Hubble parameter,

$$l_{PH} = \beta H^{-1} \quad (6)$$

Substituting Eq. (6) into Eq. (4) when Eq. (1) is used, gives

$$\rho_X = \frac{3\alpha}{8\pi}\left(\dot{H} + \left(1 + \frac{1}{\beta}\right)H^2\right) \quad (7)$$

Setting $x = \ln a$, we can rewrite Eq. (3) as, namely

$$H^2 = \frac{8\pi}{3}(\rho_m e^{-3x} + \rho_r e^{-4x}) + \frac{\alpha}{\beta}\left((1+\beta)H^2 + \frac{\beta}{2}\frac{dH^2}{dx}\right) \quad (8)$$

With the scaled Hubble expansion rate $h = H/H_0$, the above equation becomes

$$h^2 = \Omega_{m0}e^{-3x} + \Omega_{r0}e^{-4x} + \frac{\alpha}{\beta}\left((1+\beta)h^2 + \beta h\frac{dh}{dx}\right) \quad (9)$$

where $\Omega_r$ and $\Omega_m$ are the relative density of the radiation and non-relativistic matter in the present Universe, respectively. Solving Eq. (5), one obtains

$$h^2 = \Omega_{m0}e^{-3x} + \Omega_{r0}e^{-4x} + c_0 e^{-\frac{2((\alpha-1)\beta+\alpha)}{\alpha\beta}x} - \frac{\alpha(\beta-2)\Omega_{m0}}{(\alpha+2)\beta-2\alpha}e^{-3x} - \frac{\alpha(\beta-1)\Omega_{r0}}{(\alpha+1)\beta-\alpha}e^{-4x} \quad (10)$$

where $c_0$ is an integration constant.

This equation shows that we can consider the last three terms as the energy density of dark energy,

$$\rho_X = c_0 e^{-\frac{2((\alpha-1)\beta+\alpha)}{\alpha\beta}x} - \frac{\alpha(\beta-2)\Omega_{m0}}{(\alpha+2)\beta-2\alpha}e^{-3x} - \frac{\alpha(\beta-1)\Omega_{r0}}{(\alpha+1)\beta-\alpha}e^{-4x} \quad (11)$$

thus, dark energy includes three terms, the first one is slowly increasing with decreasing redshift for

$$0 < \alpha < \frac{\beta}{1+\beta} \quad (12)$$

while another is proportional to the evolution of non-relativistic matter $\sim e^{-3x}$ and radiation $\sim e^{-4x}$. The relation (12) indicates both constants $\alpha$ and $\beta$ are positive while the second one is a necessary condition to have an expanded Universe.

For conserved energy current, we obtain the pressure of dark energy as

$$p_X = -\rho_X - \frac{1}{3}\frac{d\rho_X}{dx} = -\frac{\alpha}{3}\frac{(\beta-1)\Omega_{r0}}{(\alpha+1)\beta-\alpha}e^{-4x} - \frac{((\alpha+2)\beta-2\alpha)c_0}{3\alpha\beta}e^{-\frac{2((\alpha-1)\beta+\alpha)}{\alpha\beta}x} \quad (13)$$

To have a compact and better form, it is convenient to recast Eqs. (12) and (13) as follows

$$\rho_X = c_0 e^{-\frac{2(\alpha\beta+\xi-\zeta)}{\alpha\beta}x} - \frac{\alpha(\beta-2)\Omega_{m0}}{\zeta}e^{-3x} - \frac{\alpha(\beta-1)\Omega_{r0}}{\xi}e^{-4x} \quad (14)$$

$$p_X = -\frac{\alpha(\beta-1)\Omega_{r0}}{3\xi}e^{-4x} - \frac{\zeta c_0}{3\alpha\beta}e^{-\frac{2(\alpha\beta+\xi-\zeta)}{\alpha\beta}x} \quad (15)$$

where we define

$$\zeta \equiv (\alpha+2)\beta - 2\alpha \quad (16)$$

$$\xi \equiv (\alpha+1)\beta - \alpha \quad (17)$$

The equation of state $\omega_X = p_X/\rho_X$ of dark energy is given by

$$\omega_X = -1 + \frac{\zeta}{3\alpha\beta}\left(\frac{\alpha^2\beta(\beta-1)\Omega_{r0}e^{-4x}+\zeta\xi c_0 e^{-\frac{2(\alpha\beta+\xi-\zeta)}{\alpha\beta}x}}{\alpha\xi(\beta-2)\Omega_{m0}e^{-3x}+\alpha\zeta(\beta-1)\Omega_{r0}e^{-4x}-\zeta\xi c_0 e^{-\frac{2(\alpha\beta+\xi-\zeta)}{\alpha\beta}x}}\right) \quad (18)$$

There are three constants $\alpha$, $\beta$ and $c_0$ in the expressions of $\rho_X$ and $p_X$. With aid of current observational data, one is able to find the value of these constants. For example, we set $\omega_{X0} \approx -1$, $H_0 = 67.4$, $\Omega_{m0} \approx 0.315$ and $\Omega_{m0} \approx 8.1 \times 10^{-5}$ [17]. So from Eq. (9) in the current Universe, $x = 0$, we have:

$$c_0 = \frac{\alpha^2(\beta^2-3\beta+2)}{\zeta\xi} + \frac{(2.37\alpha+1.37)\beta^2-3.37\alpha\beta}{\zeta\xi} \quad (19)$$

Furthermore, substituting Eq. (19) into Eq. (18), gives $\alpha$ as

$$\alpha \approx \frac{1.37\beta}{1.05\beta+2} \quad (20)$$

Also, by using condition (12), we obtain

$$0 < \beta < 1.99 \quad (21)$$

We have plotted the general behavior of the equation of state of dark energy in our model in Fig. 1. As shown, the evolution of the equation of state depends only on the value of $\beta$; for values of $\beta$ near zero and less than 1, dark energy evolves faster than in cases in which $\beta > 1$ and or $\beta \to 2$. However, regardless of the value of $\beta$, dark energy in our model behaves such as a phantom field in near future, approaches to $\omega_X \approx -1$ in present time while it behaves nearly like matter with $\omega_X \approx 0.33$.

To find and constraint $\beta$ by observational data one has wide plausible approaches including using different sets of Supernovae Ia data to using some methods such as the Monte Carlo Markov chain algorithm [18]. In this study, we have used the results of the joint analysis of SNe+CMB data with the $\Lambda$CDM model suggests that the transition point occurs in $z_T = 0.52 - 0.73$ [19].

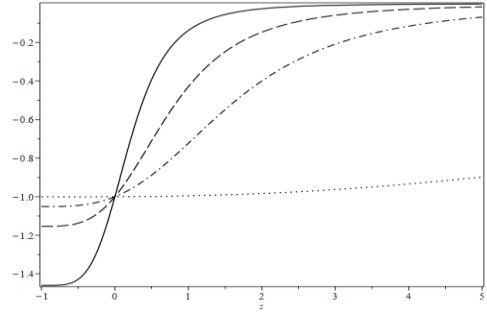

FIG. 1: The evolution of the equation of state versus redshift for $\beta = 0.5$ (solid curve), $\beta = 1$ (dashed curve), $\beta = 1.5$ (dash-dotted curve), and $\beta = 1.99$ (dotted curve).

Hence, the set of constants $\{\alpha, \beta\}$ is given by

$$\alpha \approx 0.50 - 0.66 \quad (22)$$

$$\beta \approx 1.315 - 1.99 \quad (23)$$

where Eq. (20) is used.

In Fig. 2, the evolution of the deceleration parameter of our model for three values of $\beta = 1.315, 1.657,$ and $1.99$ is plotted.

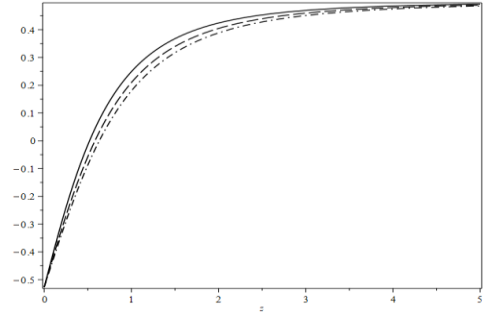

FIG. 2: The Deceleration parameter as a function of redshift for $\beta = 1.315$ (solid curve), $\beta = 1.657$ (dashed curve), and $\beta = 1.99$ (dash-dotted curve).

In Fig. 3, the age of the Universe as a function of redshift for $\beta = 1.657$ is plotted.

$$t = \frac{1}{H_0}\int_0^{\frac{1}{1+z}}\frac{dx}{h} \quad (24)$$

As shown, the age of the Universe is $\approx 13.8$ Gyr approximately which satisfies recent observations [17].

The three circles denote the ages of three old supernovae, LBDS 53W091 ($z = 1.55$, $t = 3.5$ Gyr) [20], LBDS 53W069 ($z = 1.43$, $t = 4.0$ Gyr) [21], and APM 08279+5255 ($z = 3.91$, $t = 2.1$ Gyr) [22]. It shows how our model alleviates errors of the age of these old supernovae in Ricci holographic dark energy.

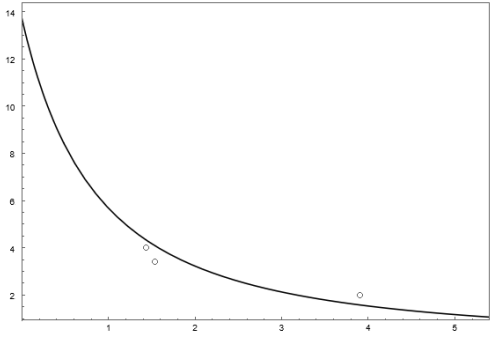

FIG. 3: Age of the Universe versus redshift for $\beta = 1.657$.

As discussed, our model gives a valuable equation of state, deceleration parameter, and age of the Universe. In the following of this study, it is worthwhile to compare our model with Ricci holographic dark energy. From Eqs. (11) and (13), energy density and pressure of dark energy in our model are, respectively

$$\rho_X \approx -15 \times 10^{-6} e^{-4x} + 0.021 e^{-3x} + 0.66 e^{0.095x} \quad (25)$$

$$p_X \approx -5.2 \times 10^{-6} e^{-4x} - 0.68 e^{0.095x} \quad (26)$$

while corresponding parameters in Ricci dark energy are

$$\rho_{X_R} \approx 0.080 e^{-3x} + 0.65 e^{0.348x} \quad (27)$$

$$p_{X_R} \approx -0.73 e^{0.348x} \quad (28)$$

where we use Planck data [17].

The general behavior of energy density and pressure of our model and their corresponding parameters in Ricci dark energy are illustrated in Figs. 4 and 5. As shown in Fig. 4, the energy density of Ricci dark energy is always bigger than the energy density in our model, in the current Universe, they are in minimum deviation from each other $\Delta\rho \approx 0.004$. The pressure of Ricci scalar model goes to zero in infinity while this parameter in our model goes to negative infinity in high redshift and so this pressure can cancel the effects of the energy density of dark energy in past. The pressure of our dark energy coincides with the corresponding parameter in Ricci dark energy in $z \approx 0.02$ while their deviation from each other is about $\sim 0.005$ in the present Universe.

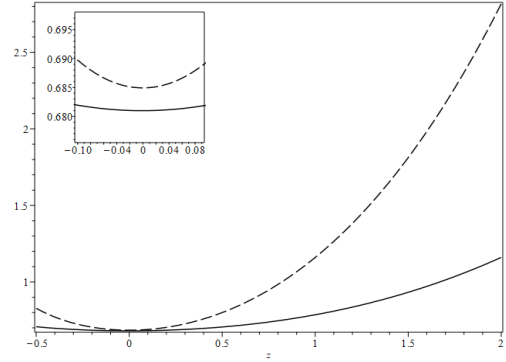

FIG. 4: The energy density of our model (solid curve) compared with the corresponding parameter in Ricci dark energy (dashed curve). We set $\beta = 1.657$ in our model.

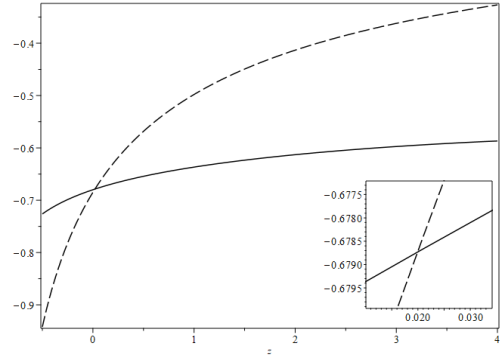

FIG. 5: The pressure of dark energy in our model (solid curve) compared with pressure in Ricci dark energy one (dashed curve) when we set $\beta = 1.657$.

In a final comparison between our model and Ricci dark energy, we have investigated the evolution of different components using modifying *COLOSSUS* code [23]. Although both models coincide with observational data in the current Universe, our model has smaller errors compared with $\Lambda$CDM theory in high redshifts, which is about $\sim 9\%$. However, more consideration demonstrates for $\beta \to 1.99$, this error in our model compared with $\Lambda$CDM theory is less than 1%. This error in Ricci dark energy model is about $\sim 23\%$.

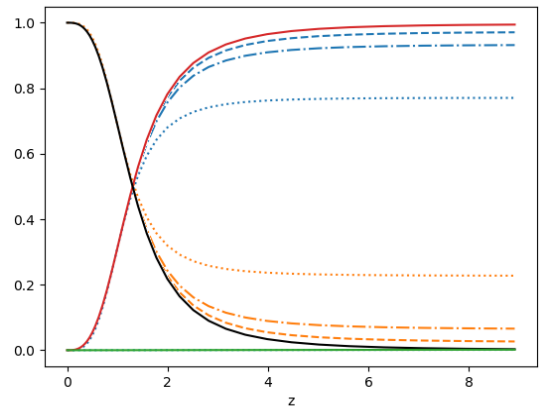

FIG. 6: The evolution of different cosmic components in our model for $\beta = 1.657$ (dash-dotted curves), $\beta = 1.99$ (dashed curves), and Ricci dark energy (dotted curves) compared with the $\Lambda$CDM model (solid curves). Blue curves represent matter evolution in our model and Ricci dark energy while orange curves present dark energy evolution in these models.

Till now, we have explored our model and compared its results with Ricci dark energy one. In the next section, we have studied structure formation in our model by using a modified version of the *CAMB* code [24].

## III. STRUCTURE FORMATION

Our model like Ricci dark energy presents dark energy that behaves such as matter in past with a positive equation of state with positive pressure. Hence, our dark energy behaves like matter during most of the matter epoch, and thus in the early matter epoch, the matter component is greater than the corresponding component in the $\Lambda$CDM model. Since dark energy in our model treats as a matter component with $\omega_X \approx 0.33$ during matter domination, the difference in matter component between our model and $\Lambda$CDM theory can be as large as about 33%. Therefore, it is maybe necessary to consider structure formation during the matter-dominated epoch and explore how dark energy in our model affects usual structure formation.

In Fig. 7, we have compared the theoretical angular spectrum of CMB temperature anisotropy of our model with the $\Lambda$CDM model. To have better insight, we calculated CMB TT, EE, and TE in Fig. 7. As shown, they differ from the standard model of cosmology mostly at small scales wherein the experimental error bars are very large and thus the model is not in conflict with current CMB observations. These largest errors in CMB TT are about $\sim 7\%$, about $\sim 14\%$ for CMB EE, and about $\sim 40\%$ for CMB TE when we fix $\beta = 1.657$. The corresponding errors are less than 1% when $\beta \to 1.99$. This demonstrates model is coincide with the $\Lambda$CDM model for the upper bound of constant $\beta$.

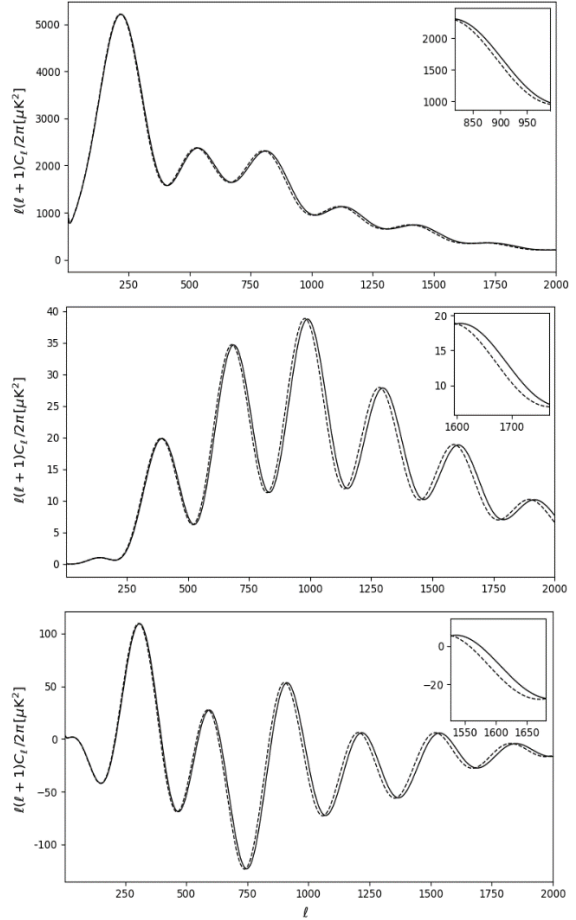

FIG. 7: The theoretical CMB TT (top panel), EE (middle panel), and TE (bottom panel) for our model (dashed curves) compared with corresponding parameters in the $\Lambda$CDM model. We set $\beta = 1.657$.

Also, we have studied the matter power spectrum of our model for $\beta = 1.657$ in Fig. 8. As expected, radiation term in energy density of dark energy alleviates dust-like component effects and so the matter-radiation equality occurred at the same $a_{eq}$ in $\Lambda$CDM model. By decreasing redshifts, the deviation between our model and the $\Lambda$CDM model grows up. However, these errors are not too large, at $k = 10^{-1}$, for $z = 0$ is less than 6%, at $z = 5, 10$ and $20$, corresponding errors are less than 2%, 0.06% and 0.006%, respectively. However, we expect these errors are no bigger than 0.001% for $\beta \to 1.99$.

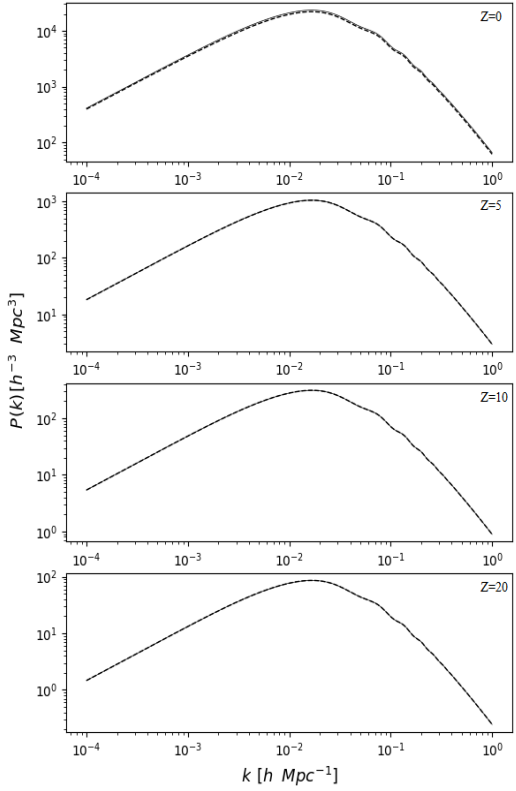

FIG. 8: The matte power spectra at different redshifts of our model (dashed curves) compared to the ΛCDM model (solid lines). We set $\beta = 1.657$.

The gravitational lensing of the CMB generates an observed polarization pattern. At the end of this study, we have explored lensing B-mode. This signal provides a measure of the projected mass distribution over the entire observable Universe and is considered a strong instrument for the measurement of primordial gravity wave signals [25]. The largest difference between our model and the ΛCDM model is about ∼4% around multipoles $\ell \approx 880$ for default accuracy. This deviation for other lens-potential accuracies is about ∼7%.

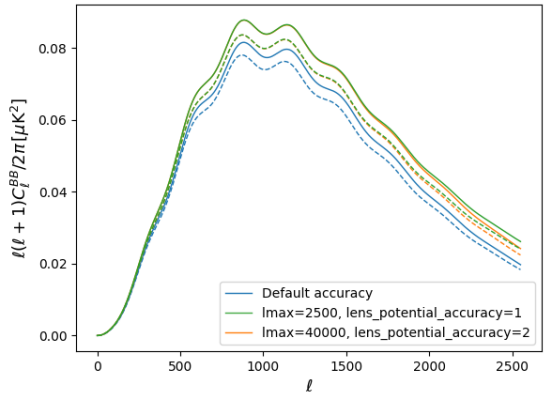

FIG. 9: The B-mode of our model (dashed curves) compared with the ΛCDM model (solid curves) for three lens-potential accuracies.

## IV. REMARKS

Dark energy as a dominant component in the late-time Universe is one of the robust unsolved problems in modern physics. The holographic principle proposes a set of dark energy models in which the dark energy density is inversely proportional to the area of the event horizon of the Universe. In this study, we have reconsidered holographic dark energy in which the energy density of dark energy comes from the acceleration of the particle horizon. This model is so similar to Ricci dark energy with different constants. As discussed, Ricci dark energy has two strong errors compared with the ΛCDM model: first in describing the age of old supernovae (Fig. 4 in Ref. [11]) and the value of cosmic components in high redshifts (Fig. 6 in this study). These inconsistencies are alleviated in our model by fixing constant $\beta$. As shown, for $\beta \to 1.99$, the behavior of most parameters such as evolution of components in high redshifts, matter power spectrum, and theoretical angular spectrum of CMB temperature anisotropy coincide with ΛCDM theory with errors less than 0.1%.

constants in the model alleviate problems of Ricci dark energy. Furthermore, considering structure formation parameters such as the angular spectrum of CMB temperature anisotropy and linear matter power spectrum demonstrates that our model coincides with observational data with so small errors.

Like Ricci dark energy model, the causality, fine-tuning, and coincidence problems are solved in our model. Since our dark energy is determined by the locally observational particle horizon, not the future event horizon, the causality problem is solved. Furthermore, since the energy density of dark energy in the current study is not associated with high-energy physics scales such as the Planck scale, the fine-tuning problem is solved. During the matter-dominated era energy density of dark energy become comparable to the Hubble parameter, and also comparable to the size of the non-relativities matter, and thus coincidence problem in the late-time Universe is alleviated.

### Acknowledgments

This work has been supported by the RUDN University Strategic Academic Leadership Program. The author thanks V. D. Ivashchuk and A. H. Fazlollahi for considering and revising the manuscript, and also referees to review and useful comments.